\documentstyle[aasms4,12pt]{article}
\raggedright
\newcommand{\mss}{mag arcsec$^{-2}$}

\newcommand{\plm}{$\pm$ }

\newcommand{\lta}{$\leq $}
\newcommand{\gta}{$\geq $}

\newcommand{\Bmo}{\rm ${\mu _B}$(0) }

\newcommand{\alp}{$\alpha$\ }
\newcommand{\etal}{{\it et.al.}\ }
\newcommand{\degree}{$^{\circ}$ }

\newcommand{\kms}{km sec$^{-1}$\ }

\newcommand{\Msol}{$M_{\odot}$\ }

\newcommand{\Lsol}{$L_{\odot}$\ }

\input{epsf}
\input{rotate}
\begin{document}
\title{\bf Red, Gas Rich Low Surface Brightness Galaxies And
Enigmatic Deviations from the Tully-Fisher Relation}
\author{K. O'Neil}
\affil{Arecibo Observatory, HC03 Box 53995, Arecibo, PR 00612}
\affil{\it koneil@naic.edu}
\author{G. D. Bothun}
\and
\author{J. Schombert}
\affil{Physics Department, University of Oregon, Eugene, OR 97403}
\affil{\it nuts@bigmoo.uoregon.edu, js@abyss.uoregon.edu}
\begin{abstract}
Using the refurbished 305m Arecibo Gregorian Telescope, we detected
43 low surface brightness (LSB) galaxies from the catalog of O'Neil, Bothun, \&
Cornell (1997a).  The detected galaxies range from 22.0 \mss\ \lta\ \Bmo\
\lta\ 25.0 \mss, with colors ranging from the blue through the
first detection of a very red LSB galaxies (B$-$V = $-$0.7
to 1.7).   The M$_{HI}$/L$_B$ of these galaxies ranges from
0.1 \Msol/\Lsol -- 50 \Msol/\Lsol, showing this sample to range from
very gas poor to possibly the most gas rich galaxies ever detected.

One of the more intriguing results of this survey is that the galaxies with
the highest M$_{HI}$/L$_B$ correspond to some of the 
reddest (optically) galaxies in the survey, raising the question of 
why star formation has not continued in these galaxies.  Since
the average H I column density in these systems is above the threshold
for massive star formation, the lack of such may indicate that these
galaxies form some kind of ``optical core'' which trace a much more extended
distribution of neutral hydrogen.  Alternatively, a model in which
no stars more massive than two solar masses form in these systems can
explain the presence of both blue and red gas rich LSB
galaxies.   Moreover, under this model the baryonic mass fraction ($f_b$) of
LSB galaxies is the same as for galaxies of higher surface brightness, thus
perhaps escaping the dilemma proposed by McGaugh \& de Blok (1997a, 1997b) with
respect to LSB and high surface brightness galaxies defining the same Tully-Fisher relation.

A subset of the detected LSB galaxies have rotational
velocities $\geq$ 200 km s$^{-1}$ and yet are at least an order of
magnitude below L$_*$ in total luminosity.  As such, they represent 
extreme departures from the standard Tully-Fisher relation.  In fact,
our sample does not appear to have
any significant correlation between the
velocity widths and absolute magnitudes, with 
only 40\% of the galaxies falling within the 1$\sigma$ low surface brightness
galaxy Tully-Fisher relation.  
Overall, the discovery of very red,
sub-L$_*$ but very gas rich LSB galaxies in the nearby Universe has increased,
once again, the overall parameter space occupied by disk galaxies.

\end{abstract}


\section{Introduction}

Our knowledge of the local galaxy population, and by inference the total
galaxy population, has been biased towards the bright, high surface
brightness (HSB) galaxies found in our current galaxy catalogs.
Low surface brightness (LSB) galaxies, those systems with central surface
brightness lower than the natural sky brightness cannot, however, be ignored,
as they are both a significant contribution to the total mass found
in galaxies and are critical to our understanding of the
distribution of galaxy types.   Without adequate representation of the
LSB class of galaxies we can not fully uncover the range of evolutionary
paths available for galaxies to follow.  LSB galaxies
are important in a number of other contexts as well:

\begin{itemize}
\item
While the overall space density of LSB disks remains unknown,
significant numbers of galaxies with \Bmo $\geq 24.0$ \mss\ have been
detected in CCD surveys, suggesting that the trend remains fairly constant
(O'Neil \etal\ 1997b) or even rises towards fainter $\mu_0$ (Schwartzenberg
\etal\ 1995, Dalcanton \etal 1997).  As a result, there is
a strong  possibility that LSB galaxies are the major baryonic repository
in the Universe (i.e. Impey \& Bothun 1997; O'Neil \& Bothun 1999).

\item  Increasing the space density of galaxies at z = 0, in principle,
poses a feasible solution to the dilemma of the apparent excess of
faint blue galaxies at intermediate redshifts.  While there is much
to be worked out in the details of such a solution
(see Ferguson \& McGaugh 1995), the discovery of faint, red, z=0
LSB spirals by O'Neil \etal (1997b) is promising in this regard.

\item
Measurements of the LSB rotation curves by Pickering \etal (1997) and
McGaugh \& de Blok (1997a) have yielded the result that
the fundamental shape/mass density of the dark matter halo appears
to be different in LSB galaxies compared to HSB galaxies of the same
circular velocity.  Additionally,  these rotation curves have shown
that the baryonic mass fractions ($f_b$) in LSB galaxies
is, on average, a factor of 3 lower than HSB galaxies of the same
circular velocity.  These observations suggest that
LSB and HSB disks are fundamentally distinct with the distinction being
physically defined by differences in their respective halos.
\end{itemize}

Although a coherent picture of the underlying mechanism of LSB
formation and evolution has not yet been found, previous studies
have shown that most of the initial, and somewhat naive, 
 concepts of LSB galaxies are at odds with the observations.
For example, the idea that LSB galaxies are ``faded'' versions of HSB
galaxies due to the dimming of an aging stellar population is shown by the
optical colors, metallicity, and gas content of LSB galaxies to be
completely wrong (Bell, \etal 1998; McGaugh 1994; Schombert, \etal 1990;
van der Hulst, \etal 1993).  
In addition, the measured low H I surface densities (van der Hulst
\etal 1993) in LSB galaxies
raise the important question:  Why are they any stars in these
systems at all (i.e. O'Neil \etal 1998; McGaugh \& de Blok 1997b;
O'Neil, Verheijen \& McGaugh 1999)?  

A recent survey by
O'Neil, \etal (1997a,b) has extended the known range of properties
of LSB galaxies, including the discovery of an intriguing new class
of fairly red LSB disks whose overall color is, in fact,
quite consistent with a conventional fading scenario.  
As this survey contains LSB galaxies with a wide range of properties,
i.e. colors from the very blue through the very red, scales lengths ranging from
1-4 kpc, central surface brightnesses from near the Freeman
value (\Bmo= 22.0 \mss) through \Bmo= 25.0 \mss, further study of
this catalog presents an excellent opportunity for understanding 
the global properties of LSB systems.  In particular, the complete properties
of the LSB galaxies discovered by O'Neil \etal (1997a,b) could not be ascertained
as no redshift measurements existed.  To further elucidate the nature of
this new LSB sample, we
undertook to detect the entire catalog of O'Neil, \etal (1997a,b)
with the refurbished Arecibo Gregorian telescope.

The layout of this paper is as follows: In section 2 the observations, data
reduction, and analysis is discussed.  Section 3 describes the 
data analysis and presents the basic observational parameters which
can be derived from the radio and optical observations.
Section 4 briefly describes some new inferences on the star formation
history of LSB spirals that can now be made with the discovery of
red, but very gas rich, LSB galaxies.  Finally, section 5 discusses how the
types of LSB galaxies discussed here do not readily conform to the
standard Tully-Fisher relation as defined by samples of cluster spirals.
Concluding remarks are contained in section 6.

\section{Observations and Data Reduction}

\subsection{Observations}

Using the refurbished Arecibo Gregorian telescope we attempted detection of  the
complete list of LSB galaxies in the catalog of O'Neil, Bothun, \& Cornell (1997a) 
(OBC from now on) available
to the Arecibo sky.  Data was taken using the L-wide and L-narrow receivers, both with 2
polarizations, 1024 channel sub-correlators, and 9 channel sampling.  Each of the four
sub-correlators has a 25 MHz bandpass, resulting in 5.2 km/s resolution (at 1420 MHz). 
The L-narrow receiver ranged from 1370 MHz -- 1432.5 MHz and 
was centered at 1401.25 MHz, providing an overlap of two correlators across 
most of the spectrum and thereby effectively doubling the integration time.
The L-wide receiver was centered at 1392.5 MHz with a range from 1350 MHz -- 1435 MHz,
resulting in only a 5 MHz overlap on each side.
In both cases the overlap more than eliminated any problems which otherwise may have 
arisen due to poor performance in the outer 50 channels of each correlator.
During the observations the L-narrow receiver was set to linear polarizations
(A or B) while the L-wide was set to circular polarization (A \plm {\it i}B).

The data were taken between 24 June 1998 and 27 April 1999. A minimum of one 5 minute
ON/OFF pair was taken of each galaxy, followed by a 10 second ON/OFF calibration pair.
When possible, if a galaxy remained undetected after the first 5 minute pair
and that galaxy was either  particularly interesting (very low \Bmo or
high B$-$V), and/or looked like a potential detection, at least one (and
on occasion many) more 5 minute ON/OFF pair was taken of it.  Table~\ref{tab:obs} lists the
total number of observations done on each galaxy, as follows:  Column 1 lists the
galaxy name; Columns 2 and 3 give the right ascension and declination (in B1950
coordinates) of the galaxies as determined from the digital sky survey (DSS);
Columns 4 and 5 list the number of 5 minute pair taken with each receiver.  If the 
pairs were of a different length of time (i.e. 2 minutes), that time is given in parenthesis;
Column 6 lists whether or not a galaxy was detected.

\subsection{Galaxy Identification}

All data was analyzed using the ANALYZ software package (Deich 1990). 
The two polarizations, as well
as any overlapping channels were initially combined to determine whether or not a 
detection had occurred.
Once an initial detection was made, each polarization and (when the detection lay
within more than on sub-correlator) each sub-correlator was analyzed to insure the galaxy 
could be retroactively detected within each system, providing a minimum of 2 -- 4 semi-independent 
detections and minimizing the chance of false positives (i.e. RFI noise).

After a detection was confirmed, the two polarizations (3-channel boxcar and hanning smoothed)
of each observation of that
galaxy were combined, and the resultant data displayed against correlator channel.
To avoid possible problems resulting from  a gain loss in the outer 50 channels of each
sub-correlator,  any detection which lay within 100 channels of a sub-correlator edge
was not used, as there was always enough overlap
in the sub-correlators  for at least one `clean' galaxy image.
Otherwise each sub-correlator detection was analyzed individually and the results were averaged.
In the cases that more than one observation of a galaxy existed, and the 
detection in question had a low S/N, the multiple observations were averaged
before the above process was applied.  If more than one observation of the 
galaxy under study was available yet the S/N was high enough within only
one ON/OFF pair for reliable analysis, each observation was analyzed 
separately with the averaged results being recorded in Table~\ref{tab:data}.

\subsection{Gain/Zenith Angle Correction}

Corrections had to be applied to the data to accommodate variations in the
gain and system temperature of the telescope with zenith angle.
The correction was determined through tracking a number
of previously studied continuum sources the entire time they were visible to the
Arecibo sky (za $<$ 19.6\degree).  The obtained fractional
temperatures (${ON\over OFF}\:-\:1$) were fit against the known
values, as published in the NRAO VLA Sky Survey (Condon, \etal 1998).
Eight  sources were observed with the L-narrow receiver and seven 
with the L-wide receiver between 
01 June through 14 June, 1998.  Fourth-order best fit curves were found
for this data and the gain and temperature variance, with zenith angle,
was obtained for each receiver.
 
To insure any observed difference in G/T between the L-wide and L-narrow receivers
was systematic,
five of the continuum sources observed with the L-wide receiver were subsequently
observed with the L-narrow, and the ratio of fractional temperatures
versus zenith angle were determined.  The result is a linear plot which shows
slight systematic variation with zenith angle, L-wide having the lowest sensitivity
at low zenith angles.  The average of the plot for all five continuum sources
is ${L-wide \over L-narrow}$ = 0.840 \plm 0.002 (3.0\degree $<$ za $<$ 18\degree).
 
As a final check, we interspersed observations of a subset of
galaxies from the catalog of Lewis, Helou, \& Salpeter (1985) with the LSB galaxy survey
described in this paper.  On average our L-narrow results differed
from those of  Lewis, Helou, \& Salpeter by 2.9\plm 0.7\%, Those
with the range in differences being $-$2\% to 10\% of their value.   The
L-wide was comparable, with an average offset of 7.1\plm 1.5\% and a range of $-$12\% -- 18\%.
See Lewis, \etal (1999) for further details.

\subsection{Data Reduction}

Baselines were fit to the data using a modified version of the GALPAK BASE module, which 
fits a polynomial to all the data within a specified region.  For this
data we let the polynomial vary in order from 1--5 for the region lying within 100--150
channels on either side of the galaxy edges.  The most sensible best fit line
(typically of order 1) was then used as the baseline subtraction.  Again,
that region within 100 channels of a sub-correlator edge was avoided for this 
analysis.

Once the baselines were subtracted, the velocities were corrected to the
heliocentric coordinate system and the velocity, velocity widths, and the
flux of the H I profiles were determined.  In each case, the values were
found four ways -- at 20\% of the mean profile flux, at 50\% of the
mean profile flux, at 20\% of the peak flux, and at 50\% of the peak flux.
(As a fraction of the detected galaxies do not show a two-horned profile,
no attempt was made to incorporate horns into the data analysis.)
During the flux analysis,
galaxy edges were estimated by eye.  Although this introduces some 
measurement error, the error lies well below the 5 -- 10 km/s width error
which exists due to the somewhat low S/N  in the 21-cm data.

\subsection{Error Analysis}

Error analysis was done as follows -- as 
no difference existed between the fluxes determined using the four
methods listed above, the determined flux errors are completely
derived from the analysis of HSB galaxies in the catalog of Lewis, Helou, \&
Salpeter (1985).
Thus the L-narrow fluxes are given errors of \plm 5\% while the 
fluxes determined from L-wide data have 10\% errors.  It is important
to note that, as determined from the Lewis, Helou, \&
Salpeter (1985) galaxies,
it is possible that the flux errors are systematic, and result in 
our presenting too low a value for M$_{HI}$.  

Comparing our central velocity values with those of  Lewis, Helou, \&
Salpeter (1985) shows virtually no
error (the highest error being 1\%), showing there is no significant contribution
from systematic errors.  The low S/N and lack of symmetry for the galaxies in this
survey, though, did result in a typical scatter between the four different
velocity measurements of \plm 2 km s$^{-2}$, on average.  If the error was larger, it is
noted in Table~\ref{tab:data}.

The low S/N and asymmetric profiles of the LSB galaxies in this
survey also resulted in considerable scatter in the velocity width measurements.
Comparing our results with Lewis, Helou, \& Salpeter (1985) suggests
our data may have systematically wider profiles by 
1 -- 2 \%.  Additionally, there was a typical scatter between the
two width determinations (peak and mean) for each fraction (20\% and 50\%) of 3\%, 
resulting in a total 5\% error for the widths (again unless otherwise
noted in Table~\ref{tab:data}).  As discussed in Bothun \& Mould (1987),
errors in 21-cm line width measurements are likely the dominant source
of observational scatter in the Tully-Fisher relation.  This problem is
made worse if most of the H I profiles do not have very steep sides.  
A quick perusal of Figure 1 immediately suggests that line width errors
are larger than ``normal'' for this sample.

\section{The Detected Galaxies}

\subsection{The Data}

The information obtained from the observations described herein are
listed in Table~\ref{tab:data} and are arranged as follows.  All 21-cm
profiles are shown in Figure~\ref{fig:HIprof}.

{\bf Column 1:}
The galaxy name as it appears in OBC.

{\bf Column 2:} 
The morphological type of the galaxy, as determined from the optical images of 
OBC.

{\bf Column 3:}
The heliocentric velocity of the galaxy in km s$^{-1}$, determined as described above.
The listed velocities have errors of \plm 5 km s$^{-1}$, unless otherwise noted.

{\bf Column 4:}
The apparent (Johnson) B magnitude for the galaxies, as given in OBC.
The error, also as given in OBC, is \plm 0.1.

{\bf Column 5:}
Absolute B magnitude, determined using 
\begin{equation} M\:-\:m\:=\:-25\:+\:5log(H_0)\:-\:5log(cz)\:-\:1.086(1\:-\:q_0)z
\label{eqn:absM}\end{equation}
(Weinberg 1972).  For this, and all other derivations in this paper, 
H$_0$=75 km s$^{-1}$ Mpc$^{-1}$, c=2.99793 x 10$^5$ km s$^{-1}$, and  q$_0$=0.1.
No k-correction was applied is it would be less than 0.1 mag.

{\bf Columns 6 \& 7:}
The average uncorrected velocity widths at 20\% and 50\% of the peak and mean  flux.
Errors are \plm 10  km s$^{-1}$, unless otherwise noted.

{\bf Column 8:}
The total flux, after correction for gain and temperature variances with
zenith angle.  No correction for partial resolution was made although
from the asymmetric appearance of some of the profiles (e.g. P3-3, P5-4)
it is possible there was some degree of poor pointing.

{\bf Column 9:}
The total H I mass, in 10$^8$\Msol, as found from 
\begin{equation} M_{HI}\:=\:2.356 \times 10^5 \: D^2\:\int{} \:S_\nu\:d\nu\:\:\:M_{\odot}
\label{eqn:MHI}\end{equation}
where the distance is in Mpc and the flux ($S_\nu$) is in Jy  km s$^{-1}$.

{\bf Column 10:}
The H I mass to luminosity ratio in units of \Msol/\Lsol, where the luminosity 
was determined from the magnitude listed in column 4.

{\bf Column 11:}
The H I mass to luminosity ratio in units of \Msol/\Lsol.  In this case the luminosity
was determined through the total integrated blue magnitude (m(\alp)) given in OBC and
converted to absolute magnitude using Equation~\ref{eqn:absM}.

{\bf Column 12:}
The B$-$V color as given in OBC.

{\bf Column 13;}
The galaxy's optical inclination, determined from the OBC data using the IRAF ELLIPSE
parameter.  The inclination listed is simply $i\:=\:sin^{-1}(r_{minor}/r_{major})$
The inclination error is \plm 5\degree unless otherwise noted.

\subsection{The Non-Detected Galaxies}

Table~\ref{tab:nondet} lists all of the galaxies for which the attempts at
detection failed during this survey.  Column 1 of the table gives the
galaxy name, while Columns 2 and 3 provide the 1$\sigma$ detection limits
for both the L-narrow and the L-wide set-ups, found from the r.m.s.
error in the baselines.  It is likely that were a galaxy
to to lie at least 1.5$\sigma$ above the baseline it would have been detected.
It is important to note, however, that this simply limits the maximum H I
density of the undetected galaxies {\it if they lay within the 0 --
11,500 km s$^{-1}$ survey boundary.}  As many galaxies were found near the
11,500 km s$^{-1}$ survey boundary, it is quite possible that a planned
search at higher redshift will readily detect many of the `missing' galaxies.

\subsection{Galaxy Distribution}

The distribution of the galaxies in the survey follows the large-scale HSB
galaxy distribution.  Figure~\ref{fig:rav} shows a cone diagram for all the 
galaxies in Pegasus and Cancer fields of OBC with known velocities (obtained 
either through this survey or through NED, the NASA Extragalactic Database).
As can be readily seen, all the galaxies lie within the HSB defined galaxy clusters,
and none of the detected galaxies fill in the cluster
voids.   
 The one exception
is U1-4 (not shown) which appears to be a genuine isolated system. That 
none of the detected galaxies fill in the cluster voids is in agreement with
both the result of Schombert \etal (1995) for LSB galaxies off the Second
Palomar Sky Survey and the Uppsala General Catalog search by Bothun, 
\etal (1986).  The reader should also note that these galaxy
clusters do not match the extreme regions of galaxy density, such as the
Coma or Virgo clusters.  We make no statement on existence of LSB galaxies
in cluster cores, already shown to be rare by Bothun \etal (1993).

A follow-up study into the clustering statistics of all these galaxies
(O'Neil \& Brandenburg 1999) should clarify this issue.  For now it suffices to
note that no trend was found between a galaxy's velocity or cluster
identity and its color, surface brightness, rotational velocity,
inclination, and absolute magnitude.  It is interesting to note that with
the inclusion of our LSB galaxies, galaxy groups now found to contain the
full range of galaxy types, from dwarfs to intrinsically luminous LSB
galaxies, and from the very blue, star-forming galaxies to those galaxies
with apparently old stellar populations.

\subsection{Galaxy Photometry, Morphology and Structure}

The galaxies for the sample span a range of morphological types from
late-type spiral to dwarf irregulars.  The Hubble types are given in Table~\ref{tab:data},
following the prescription of Schombert, Pildis, \& Eder (1997) and Sandage \&
Binggeli (1984).  This system is based mostly on the presence, and size of
a bulge component.  Objects with a distinct bulge are classed Sb to Sc
depending on bulge dominance.  Objects with a disk-like shape and some
central concentration are classed as Sm.  Objects with no shape or
concentration are classed as Im.  Most of the objects with double-horned
H I profiles, indicative of rotation, displayed disk-like morphological
appearance. Eight of the 43 objects (19\%) are classed as Im or dwarf-like.
All of these dwarf galaxies have luminosities and sizes on the low end of the
sample scale, justifying their classification as dwarf in terms of size and
magnitude.  Several of the disk galaxies in the sample have magnitudes and
scale lengths comparable to the class of dwarf spirals (Schombert \etal
1995), but many also have rotational masses similar to the most massive disk
galaxies known (see Section 5).   This immediately suggests detection
of objects that strongly violate the Tully-Fisher relation.

Figures~\ref{fig:Bcomp} and \ref{fig:morph} compare M$_B$ with the
total neutral hydrogen mass, color, central surface brightness, and
scale length for all the galaxies detected with our survey.
In Figure~\ref{fig:Bcomp} these results are compared
with other samples (de Blok, van der Hulst, \& Bothun 1995;
de Blok, van der Hulst, \& McGaugh 1996; Schombert, \etal 1995;
Matthews, \& Gallagher 1997; Bothun, Sullivan, \& Schommer 1982; Becker, \etal 1988). 
The most relevant comparison might be with
Bothun (1982) who reported the results for normal spirals in
the Pegasus and Cancer regions.  With respect to Figure~3a,
our sample clearly defines the same locus of points with the obvious
exception of 3 outliers --  C1-2, C5-5, and N9-2 (having extremely
high M$_{HI}$/L$_B$ ratios).
However, most of the points sit above the
ridge line fit to these samples, indicating that, on average, our
sample galaxies have higher gas to star ratios.  The situation summarized
in Figure~3b however is quite different.  Here we can
see that our sample has clearly extended the color-magnitude relation
for spiral galaxies with gas in that we have detected a number of
red galaxies with luminosities below that of L$_*$.  Previous reports of
very red gas-rich disk galaxies (e.g. Schommer \& Bothun 1983, 
van der Hulst \etal 1987) have all been relatively luminous spirals.
To our knowledge, this survey is the first to have discovered red,
gas-rich objects of relatively low intrinsic luminosity.  Overall,
however, there clearly is no color-magnitude relation defined for the
late type samples shown here.  These objects clearly define a very
diverse galaxy population.

In Figure~\ref{fig:morph}{\it a} we plot the absolute B magnitude versus
\Bmo for the galaxies in this survey.  By selection, there are no
galaxies with \Bmo brighter than 22.0 \mss. 
The void in the top right corner is most likely artificial, 
created by the inability of OBC to identify small, compact galaxies
(the `star-like galaxies' of Arp 1965).  Had high enough resolution
images been taken by OBC, that corner might disappear.  Indeed,
that corner is often inhabited by blue compact dwarfs which have
become the source of intense recent study (e.g. Marlowe \etal 1999).
The more interesting void in Figure~\ref{fig:morph}{\it a} is at the lower
left corner.  This is the region where very LSB yet intrinsically
very luminous, and physically quite large, galaxies would lie (the Malin I galaxies).  This
lack of intrinsically luminous, very LSB galaxies can also be seen 
in Figure~~\ref{fig:morph}{\it b}, which shows a void in the very LSB, large
scale length regime.   This result either shows that the intrinsic
space density of these physically large LSB disks is low (but see Sprayberry \etal
1995a) or that we have not sampled the correct regions to find them;
that is, they tend to avoid the extended
cluster environment (see Hoffman, Silk, \& Wyse
1992) are are only found in isolated, very low density environments.

\section{Red Gas Rich LSB Galaxies}

The use of the color-gas content plane as a diagnostic for studying
the evolution/star formation history of galaxies was pioneered
by Tinsley (e.g. Tinsley 1972).  Bothun (1982) present 
a codified version of this as it applies to cluster spirals.  One
of the main results of that analysis was that, for a sufficiently
large sample of spirals, there really was no correlation between disk
color and the gas-to-star ratio.
Figures~\ref{fig:red} {\it a} and {\it b} show the
H I mass and M$_{HI}$/L$_B$ versus B $-$ V color for the galaxies in our 
survey  where we confirm the absence of any correlation.  Note the
presence of some galaxies in our sample which have extreme values of
M$_{HI}$/L$_B$.
 More to the
point, our samples clearly shows that there is  
no definitive trend toward lower M$_{HI}$/L$_B$ or M$_{HI}$
with redder colors.
In fact, if a trend does exist, it is toward {\it higher} H I mass and 
M$_{HI}$/L with increasing B$-$V, raising an interesting question -- 
if a galaxy has been capable of forming and evolving at least one generation
of stars to reach
B$-$V \gta 1.0, and the galaxy still contains a significant fraction
of neutral hydrogen, why has it been unable to 
continue this process?  That is, why hasn't such a disk exhausted its
gas and assumed its endpoint evolution in this diagram at the lower right
where the bulk of S0 galaxies reside?  Clearly, we have detected what
could best be described as a ``dormant'' population of disk galaxies.

One solution to the above questions may be found by appealing to Kennicutt's (1989)
argument that a critical density of gas must be present for star formation to
occur.  Thus it could be argued that star formation in these red, high 
M$_{HI}$/L$_B$ galaxies occurred only in those regions dense enough to
support it, and that the bulk of the gas is located in the less dense regions which have been unable to commence self-gravitation.    In this case,
these galaxies would represent ``optical'' cores inside a much more extended
gas distribution.  Certainly individual examples of this have been 
found (e.g. NGC 2915 -- Meurer \etal 1996), but these seem to be quite
rare.  Indeed, a systematic search for such kinds of objects from a 
seed sample of blue compact dwarfs failed to detect any examples of
large H I mass objects with small optical sizes (i.e. Salzer, \etal 1993).  

Previous studies of red, H I rich disk galaxies (e.g. van der Hulst
\etal 1987) have shown them to be large scale length objects with
generally strong color gradients.   Their overall properties imply
that the bulk of the star formation has occurred within the inner
two scale lengths and the gas at larger radii is mostly unprocessed.
The red H I rich galaxies in this survey, however, are not of large
scale length as shown in  Figure~\ref{fig:MLBalp}.
In fact, for the extreme cases, it can be seen that the galaxies 
with M$_{HI}$/L$_B$ \gta\ 9 \Msol/\Lsol are confined to 
the \alp $<$ 2 kpc range.  Possibly this is an indication that
these galaxies are indeed ``optical'' cores of a much larger H I extent.
However, if their H I size is similar to their optical size then we
have discovered a class of relatively compact galaxies which have an
average column density of approximately 10$^9$ \Msol/100 kpc$^2$ =
10$^{21}$ atoms per cm$^{-2}$ (e.g. above the threshold), but which
are not forming stars, or at least are not forming massive stars.
It is this latter possibility that makes the discovery of these
objects relevant 
to the apparent contradiction discussed by McGaugh \& de Blok (1997a) regarding
the low $f_b$ value that results for LSB galaxies if they are fit with
a standard dark matter halo.   

This low $f_b$ value is follows directly from
assuming a low M/L for the stellar population.  Typically
a value of (M/L)$_B$ of 1--2 is assumed as this is consistent with 
the relatively blue colors of the McGaugh \& de Blok disks.    But
what if this is wrong?  What if we are dealing with an anomalous
stellar population because of an anomalous IMF?  Suppose, for the sake
of argument, that we force the $f_b$ to be the same for both LSB
and HSB galaxies.  One motivation
for doing this is to better understand why it is that LSB disk galaxies
apparently define the same Tully-Fisher (TF) relation as HSB disk galaxies (e.g.
Sprayberry \etal 1995b; Zwaan \etal 1997).   This issue has been
addressed by Dalcanton \etal (1997), Mo, Mao, \& White (1998) and 
Steinmetz \& Navarro \etal (1999) who can recover this ``universal''
TF relation by assuming that galactic disks are sub-maximal and it
is the halo properties that therefore dominate the mass profile.
Support for this general picture comes from a recent study by Courteau
\& Rix (1999) who show that the TF relation and its scatter are best
understood if the disks of HSB spirals are indeed sub-maximal.  On
the other hand, Beauvais \& Bothun (2000) present fairly compelling
evidence that disks are likely maximal (see also Sackett
1997; Giraud 1998).

Mo, Mao, \& White (1998) explore the efficiency of the conversion of baryonic
mass into stars.  Variations in this efficiency can obviously lead
to different loci of points in the TF relation.  In the extreme case
of no conversion into stars, obviously a galaxy has no luminosity
to even place it on the TF relation!  Mo, Mao, \& White (1998) find that
surface mass density is a key parameter in determining this efficiency
(see also Bothun 1990) and predict that low surface mass density
systems have low efficiencies leading to the existence of relatively
gas-rich but dim galaxies.  Such galaxies, are in fact, the kind that
we have discovered here.  

However, the luminosity evolution of galaxies depends on factors other
than the efficiency of gas to star conversion.  There is the critical
issue of the mass function of stars that are formed during the
conversion process.  While discussing whether or not there is a universal
IMF is well beyond the scope of this paper,
we amplify the issue that was initially raised in McGaugh and
de Blok (1998) -- at face value, the data seem to argue that
two potentials of much different baryonic
mass fraction, but similar circular velocity, produce the same
amount of light (hence making a universal TF relation).  While this
issue can be resolved using Modified Newtonian Dynamics 
(MOND e.g. Milgrom 1983), that
solution seems no more extreme than a solution which seeks to change
the stellar population so that the baryonic mass fractions in LSB
galaxies in fact, is the same as in HSB galaxies.

For $f_b$ to be
the same, the assumed (M/L)$_B$ for blue, LSB disks would have to be
10, instead of  1--2.   While there may be some dynamical constraints
of having  (M/L)$_B$ this high (Quillen \& Pickering 1997), values
this large cannot be explicitly ruled out for the objects in the
de Blok and McGaugh (1998) sample.  They are only indirectly ruled out
by the standard (M/L$_B$)/$B-V$ arguments which flow from an
assumed IMF.  Therefore, one is obligated to produce a stellar
population which basically has (M/L)$_B$ $\sim$ 10 (the value for a typical
elliptical galaxy) and $B - V$ $\sim$ 0.3.  This is a tough problem
and its solution requires an extreme stellar population, one that is
presently deficient in stars on the giant branch,
so that the high (M/L)$_B$ is
provided by main sequence stars of mass less than 0.3 \Msol and the
blue colors are supplied by a small amount of A and F stars.  This
population of A and F stars
must remain sufficiently small so that their subsequent 
giant branch evolution does not significantly alter (M/L)$_B$. 
Note that a galaxy with such a stellar population will not remain
dim forever, only for the first few billion years of its life. 
After 10 Gyr, the heavily populated lower main sequence will begin
to evolve onto the giant branch and the luminosity will increase.

This extreme model, in fact, is observationally testable if near IR
colors could be obtained.  Dwarf dominated integrated 
light will be significantly bluer than giant dominated integrated light
in $V-K$.  We also note that it is very difficult to
produce any stellar population with giant branch dominated light
that has $B-V$ $\geq$ 1.0 (see Tinsley 1978; Bothun 1982) and ages
that don't exceed 12 Gyrs.  The reddest objects in our sample 
approach $B-V$  = 1.2.  Since there is unlikely
to be significant internal extinction in these systems (see
below), these very
red colors imply either 1) a rather old galaxy (unlikely) or 2)
the presence of very cool giants usually only seen in metal-rich 
populations (again unlikely), or 3) a large contribution to the light
from main sequence stars of spectral type K5 -- M0.

In the galaxies' star forming (ON) state it is blue.
In the galaxies non-starforming (OFF) state  it fades and reddens.   
In Table~\ref{tab:botmod}, we present three models (discussed in 
detail below)
which show the ON and OFF states  that can be achieved under our
dwarf dominated integrated light scenario.  While this scenario
is clearly non-standard it does produce both red and blue gas-rich
LSB galaxies with high values of $M/L$ in both cases.  As such,
this scenario provides an alternative to apparent significant discrepancy
between the different $f_b$ values inferred for LSB and HSB disks.
However, to be fair, in reaching this conclusion,  McGaugh \& de Blok (1997a,b) 
actually  compare the $f_b$ of LSB disks with clusters of galaxies,
which typically have $f_b$ in the range 0.1-0.3 (e.g. the baryon
catastrophe -- see White \etal 1993). By contrast, LSB disks are
factor of 3-5 lower.  Direct comparison to HSB disks, however, is
more problematical as most rotation curves don't go out far enough
to accurately sample the halo.  Thus, $f_b$ depends upon the scale
of measurement in the HSB sample.   Adding to the confusion are the
results of Zaritsky \etal (1997) whose work on satellite galaxies
and the total extent of the halos around HSB galaxies leads to
$f_b$ ~0.05, similar to those of LSB disks but significantly less
than what is seen in clusters.  So while its clear that $f_b$ for
LSB disks is significantly lower than that measured for clusters 
of galaxies, it remains ambiguous whether or not $f_b$ is also
significantly lower compared to individual HSB disks.

This lack of clarity on the $f_b$ issue, however, does not deter us
from pursuing our alternative stellar population model.  Indeed,
in a rigorous but now largely forgotten paper, Larson (1986) makes
a set of compelling arguments about the need for a bi-modal IMF 
to fully reconcile the chemical evolution of galaxies with their
observed colors and mass-to-light ratios.   
One of those modes involves a peak at M $\sim$
0.25 \Msol and that is the mode we have adopted as the dominant
mode of star formation in our models.   To account for the presence
of very blue LSB galaxies and the observation (e.g. McGaugh 1994)
that some LSB galaxies have a small number (e.g. 1--6) of H II regions 
we have added a few O,A and F stars to this mode.   However, as pointed
out in de Blok (1997) as well as O'Neil \etal (1998), there is ample
evidence that LSB disks are generally deficient in the formation
of massive stars.  The models of O'Neil \etal (1998) assert that this
deficiency is a statistical result of a low surface density of gas
and the difficulty of gathering sufficient gas mass over a relatively
small spatial scale to produce the kind of massive star formation that
is observed in HSB disks.  

Under this set of assumptions we ran a large series of models and
show the results for 3 specific cases in Table~\ref{tab:botmod}.
In each case, the ON state is achieved by allowing $\sim$ 2\% of the
available gas mass to to participate in star formation.  Values for
$B-V$ and $M/L$ for the stellar population are listed for the ON
and OFF states.  The extreme nature of these models is readily apparent
as the giant branch contribution to the integrated light at B is never larger
than 10\%.   In models 1 and 2 we truncate star formation at 2 \Msol so
that only A and F stars are produced.  The main difference between the
two models is that model 2 peaks at a lower main sequence mass to
drive the colors as red as possible.  Still, none of our models can reach
the more extreme $B-V$ values in this sample (e.g.  $B-V$ $\geq$ 1.2).
In model 3, we add O-stars to model 2 in order to accommodate 
the existence of the observed H II regions.   This addition does not
appreciable lower the $M/L$ for the stellar population but does push
the $B-V$ color blueward by $\sim$ 0.1 magnitude.   Since only a 
relatively small amount of gas is involved in the OFF to ON transition,
then our models suggest that both blue and red H I rich LSB galaxies
can co-exist for billions of years, given their amounts of available
H I.  Note as well that in all
three cases, the integrated light in the ON state is dominated by the
small number of O,F and A stars.  As the light from these stars fades,
the luminosity of the galaxy will decrease by a factor of 3-4 and
the galaxy will return to its red OFF state.   In fact, if in its ON
state the galaxy falls on the standard TF relation, it will surely
be well below this position as it transitions to the OFF state.
In the next section, we will show that these red H I rich LSBs are
significantly under-luminous with respect to their circular velocity.

We have pursued this alternative stellar population model strictly 
as means of determining
the plausibility of producing a stellar population which can be blue
($B-V$ $\sim$ 0.4) but which can have a significantly higher $M/L$ ratio
so as to raise the value of $f_b$ for LSB disks to the level observed
for clusters of galaxies.
The models presented in Table 4 achieve this and we leave it up to
the reader to assess their plausibility in comparison to other
alternatives (e.g. a non-universal value for $f_b$, MOND, etc).

Clearly, further resolution of this intriguing matter requires actual
measurements of the H I distributions in these red H I rich objects.
Either the gas is well below the critical density thus effectively shutting off
star formation, or star formation is proceeding without the accompanying
massive star formation.  While that would be very strange, we argue
that the position of these galaxies in the color-gas content plane
is already very strange so something quite different is likely going on.
The fact that we can produce blue LSB galaxies in a model in which only
2\% of the gas supply needs to participate in a star formation event
is consistent with what appears to be occurring in UGC 12695, perhaps
the bluest LSB detected to date.  UGC 12695, in fact, has a total H I mass
remarkably near its total dynamical mass ($M_{HI}\:=\: 4.2 \times 10^{9}$\Msol,
M$_{dyn}\:=\:9.6 \times 10^{9}$\Msol), yet recent HST Wide Field Planetary Camera-2 and
VLA observations show that the majority of star formation in UGC 12695 
is occurring in isolated regions.  Thus, although UGC 12695 has a high
total H I mass, the surface density of the gas is often below 
the Kennicutt star formation threshold (O'Neil,
Verheijen, \& McGaugh 1999; O'Neil, \etal 1998).  If the
light from this small amount of star formation fades, its quite likely
that UGC 12695 would be a red, LSB, H I rich object.

\section{Deviations from the Tully-Fisher Relation}

In 1977 Tully \& Fisher published an empirical correlation between
galaxies' H I velocity width and their absolute magnitude (or diameter).
Since that time the Tully-Fisher (TF) relation has been used in over 100
different investigations that attempt to determine the relative
distances between disk galaxies and/or clusters of galaxies.   This
multitude of different investigations clearly establishes that
the TF relation exists in a variety of different samples.
The recognition
that the potentials of disk galaxies are dark matter dominated makes
the existence of the TF relation very difficult to understand from
first principles because, apparently, it requires that the amount of
luminosity trace the dark matter in a way that scales exactly with $V_c$.
This fine tuning, in essence, requires that the value of $f_b$
in disk galaxy potentials be nearly constant.   The required constancy
of $f_b$ should be a troubling point.  
The existence of the TF
relation implies that just enough baryonic gas creeps into these
dark potentials and then has the correct star formation history to produce
the right amount of light for a given circular velocity.   Worse
still is the apparent physical dilemma or paradox that results from
the apparent ability for samples of LSB and HSB spirals to define
the same TF relation.  Simple virial arguments coupled with the
constancy of $M/L$ fail
to reproduce this observation if LSB and HSB disk galaxies are dominated
by dark matter halos of the same form (see Bothun \& McGaugh 1999).
The only way out of this dilemma is to postulate that fine tuning 
exists between central surface brightness and $M/L$ in such a way
so as to preserve a universal TF relation.

In this section, however, we will document how our, relatively extreme
sample of LSB galaxies, in fact does not conform well to the standard TF
relation.  While hints of this have been seen in other samples
(e.g. Matthews, van Driel, \& Gallagher 1998) those hints are mostly
based on behavior at the low line width end.  Our sample contains 
galaxies with substantial circular velocity that nevertheless strongly
deviate from the TF relation.

\subsection{Inclination, Extinction, and Random Motion Corrections}

To compare our results with those of previous galaxy surveys, we modeled
our magnitude and line width corrections
after those of Tully \& Fouque (1985) and Zwaan, \etal (1995).  Thus we
corrected the magnitudes for Galactic and internal extinction to face-on 
orientation, using
\begin{equation} B^i\:=\:B\:+\:2.5log\left( f\left( 1\:+\:e^{-\tau/cos(i)}\right)\:
+\:(1\:-\:2f)\left( {{\left( 1\:+\:e^{-\tau/cos(i)}\right)}\over{\tau/cos(i)}}\right)
\right)\end{equation}
with $\tau$=0.55 and {\it f}=0.25.
The line widths  were corrected for random motion effects and inclination, using
\begin{equation}W^i_{50_C}\:=\: \sqrt{\left({{W^i_{50}}^2 \:+\: W_t^2\:-\:2W^i_{50}W_t
\left({ 1-e^{-\left( {W^i_{50}/W_t}\right)^2} }\right)
\:-\:2W_t^2
\left({ e^{-\left({W^i_{50}/W_t}\right)^2} }\right)
}\right)} \end{equation}
with $W^i_{50}\:=\:W_{50}\left({ cos(i) }\right)^{0.22}$ and 
W$_t$, the turbulent motion parameter, being set to 14 km s$^{-1}$.

Figure~\ref{fig:tf}(a) shows W$^i_{50_C}$ versus B$^i$ with the 1$\sigma$ and
2$\sigma$ ranges of Zwaan, \etal overlaid on the plot.
The Zwaan, \etal data, based off Broeils (1992) findings, is
\begin{equation}M_{B_{corr}}\:=\:-6.59\:log(W_{50_{corr}})\:-\:3.73\:\:\pm0.77\end{equation}
Only 77\% of the data falls within the 2$\sigma$ lines of Zwaan, \etal,
and a mere 40\% fall within the 1$\sigma$ lines.  Clearly this sample
does not produce a well-defined TF relation for LSB galaxies in the
way that the sample of Zwaan \etal does.

\subsection{Potential Errors}

Before delving into the significance of Figure~\ref{fig:tf}(a), a further look into
potential errors is required.  The first question which needs to be addressed
is that of potential biases introduced by above
corrections, as all were done as if this sample was comprised
of HSB galaxies.  

The application of these ``corrections'' however, to LSB galaxies may
be somewhat dubious.  To begin with, there is a fair bit of evidence
that suggests LSB galaxies have very low amounts of internal extinction and
may, in fact, be extinction free.  For instance, Sprayberry \etal (1995b)
did not apply any extinction corrections to the magnitude and formed
a TF sample with a slope very near the virial slope of $-$10.  
The extinction correction assumes the galaxies are at least
moderately optically thick.  In Figure 8 we  plot the B$-$V vs B$-$I colors of 
the galaxies in OBC.  A nice locus of points is found whose position
matches simple expectations of stellar population models.  Where the
red galaxies driven so by excessive amounts of dust we would observe
significant shifts away from this relation.  The tightness of the
two color diagram thus provides another indication that 
LSB galaxies are relatively dust-free (Figure~\ref{fig:bvbi}).
Additionally, it should be noted that the six most deviant points in Figure~\ref{fig:tf}(a)
(P1-3, P9-4, C4-1, C5-5, C6-1, C8-3) range in color from B$-$V = 0.5 -- 1.0
and in inclination from 45\degree through 73\degree.
It is therefore likely that, if anything, the extinction corrections applied to the data
are larger than necessary, and the true B magnitudes are fainter than shown in 
Figure~\ref{fig:tf}(a).  This, then, would shift the data points downward, and farther
from the Tully-Fisher relation.

Perhaps a better gauge of the departure of our LSB sample from the standard
TF relation is provided in Figure~\ref{fig:tfi} which shows the I-band 
Tully-Fisher relation for the data in this paper, with the data and $1\sigma$ 
lines from Pierce \& Tully (1988) overdrawn.  From this it can be seen that the
scatter is as deviant in the I band as in the B for our data.
It should be noted that the two data points (U1-4, and I1-1)
which lie above Pierce \& Tully's
TF relation in Figure~\ref{fig:tfi} all belong to galaxies which are not 
a part of any known cluster and hence we can not assign some mean "cluster"
distance to these galaxies as we can for the bulk of the objects in
our sample (see Figure~2).

The next correction was done to compensate for the inclination of the
galaxies when trying to determine their rotational motion.  This correction is
fairly straightforward provided the true inclination of a galaxy is known.
The inclination of the galaxies in this survey were determined by OBC assuming 
their true (face-on) shape is a perfect circle.  LSB galaxies
tend to be more amorphous in appearance than their HSB counterparts,
however, making optical determinations of their true inclination
difficult.  By their very
nature, LSB galaxies are difficult to fit isophotes to and there is
a fair amount of variation in position angle and ellipticity between
adjacent isophotes.  While some of this noise averages out, there is
likely an increase scatter in the TF relation for this sample just
as a result of ``inclination'' noise.    It is unclear, however, if
there is any systematic effect present in applying the inclination
correction to the measured line width.  It should be noted that the
\plm 5\% error in inclination is accounted for in the error bars
of Figure~\ref{fig:tf}.

The final correction undertaken was to compensate for the random 
motion of the gas contained within each galaxy.  This is 
the most uncertain of all the corrections, as the parameter which
describes the importance of turbulent motion, W$_t$, is
unknown for LSB galaxies.   For HSB galaxies, W$_t$ appears to be
of order 10-15 km/s, averaged over the entire disk (see data in
Beauvais \& Bothun 1999).  For LSB galaxies one might expect
W$_t$ to be larger due to the lower restoring force perpendicular
to the disk (due to low surface mass density).  On the other hand,
W$_t$ also is a measure of the kinematic feedback of star formation
to the gas and LSB galaxies have quite low current star formation rates.
Thus it is possible that W$_t$ is either significantly
greater than (or smaller than) that used in the above equations,
which would shift the points in Figure~\ref{fig:tf} to the
left (or right).

Since the application of our corrections to LSB galaxies may have limited 
validity, it seems worth just plotting the data without the corrections.
This is done in Figure~\ref{fig:tf}(b) which 
shows the same galaxies
without the corrections for inclination, internal extinction,
and random motions (only the Galactic extinction correction remains).
The situation with respect to the uncorrected data, in comparison with
the lines of Zwaan, \etal (1995), is very similar to that in the corrected
data, strongly indicating that the observed (large) deviations from the
TF relation are real, and not the result of large systematic errors 
associated with applying the corrections.

\subsection{Why no Tully-Fisher Relation?}

The most obvious question to raise in regards to our results is why
our survey did, and previous surveys did not, find galaxies deviating
so widely from the Tully-Fisher relation.    One possibility is simply
the sensitivity of our survey and the size of this LSB sample.
The combination of the large number of data points and the
increased sensitivity of the Arecibo Gregorian system make for the ready detection
of low S/N galaxies (i.e. P6-1 and P1-3).  
The second part of the answer to the above question is that 
the discovery of a group of galaxies which do not follow the 
Tully-Fisher relation is not entirely unanticipated.   For instance,
it should be clear that when a galaxy is a big bag of baryonic gas
and has not produced any stars (light) yet, that it can not lie
on the TF relation.  So, one naively expects that galaxies with high
fractional gas contents should systematically be under-luminous for
their line width and hence fall-off the TF relation.  However, no
sample to date really shows this, and attempts to look for a correlation
of the residuals with gas-content (e.g. McGaugh \& de Blok 1997b) have
not produced anything convincing.  What is unique about our data set
is the appearance of galaxies with relatively large line widths that
are deviating from the TF relation in this expected sense (e.g.
Figure~9).  In general
those are the points with $W_{20_{corr}}$ $\geq$ 200 \kms.  
At the lower line width end, a number of studies have showed that
the dispersion around the TF relation starts to increase and some
galaxies are located well below the nominal relation
(e.g.  Matthews, van Driel, \& Gallagher 1998).

In general, our sample has H I profile morphologies consistent with
rotating disks (e.g. we see 2 horns).  Some low line width
galaxies show Gaussian or triangle profile morphologies.
Inclusion of such galaxies invariably increases
the scatter at the faint end because it is unclear if
there is even a kinematic disk present that is rotating at some angle
with respect to the line of site.  So one shouldn't pay to much attention
to the behavior in the TF plane for those galaxies with rotational
line widths $\leq$ 80 \kms.

In Figure~\ref{fig:resid} we plot the residuals from Figure~\ref{fig:tf}{\it a}
as a function of M$_{HI}$/L$_B$.
A definitive trend toward the more H I rich objects lying the
farthest off the Tully-Fisher relation is evident.    This
reflects a substantial reservoir
of unprocessed gas and hence these potentials are under-luminous for
their rotational velocity.  While this is completely obvious, this
is really the first sample where this effect is directly seen.  This,
of course, doesn't mean that there isn't a TF relation, it just means
that the TF relation as defined by the luminosity versus circular velocity
plane is not appropriate.  If we replace luminosity by baryonic mass
then there will be a large correction that is driven by the gas mass
fraction.  Discussion of the existence of a "baryonic" Tully-Fisher
relation, however, is beyond the scope of this paper.
However, de Blok \etal (1999) have undertaken a systematic
investigation of the baryonic TF relation to conclude a) that such
a relation exists and b) it has relatively low scatter.  Under the
idea of constant $f_b$ per potential, a good baryonic
TF relation must exist.  Galaxies with high gas mass fraction must
therefore naturally depart from the L-V relation in the sense that
is finally observed in our sample.

The previous lack of detection of galaxies lying outside the 
2 $\sigma$ range of the H I Tully-Fisher relation in Figure~\ref{fig:tf}
thus appears to be due to selection effects similar to those which created the
`Freeman law' and delayed the detection of LSB galaxies.  
Also akin to the original optical non-detection of LSB galaxies, these
selection effects have been discussed in the literature, most notably
by Briggs (1998) and Disney \& Banks (1998).  The H I Tully-Fisher
relation, then, like  the Freeman law, is a relation demonstrating
our limited view of the universe, as only galaxies with similar
gas mass fractions have been surveyed to date.  Since gas turns into stars,
the evolution in surface brightness of any disk galaxy is equivalent
to evolution in decreasing gas mass fraction.  The difficulty of finding
high gas mass fraction galaxies is therefore equivalent to the general
difficulty of detecting galaxies of extremely low surface brightness.
The fact that such galaxies are now shown to exist in the nearby universe
is once again a testimony to the slow evolutionary rates of some
galaxies (e.g. Bothun \etal 1990).  Were we to have done an all sky
survey at 21-cm before any optical surveys its likely that a) 
Freeman's Law would have never existed and b) the Tully-Fisher relation
would have been harder to find.
As more observations are done, and as the 
sensitivity of telescopes increase, we predict that most
all possible regions of the H I velocity width versus 
magnitude plot should become populated.  Indeed, one can imagine 
populating the upper left region simply as a result of a small starburst
in one of these dormant galaxies.  The sample of Salzer \etal (1999)
may contain such objects, but of course, such objects are {\bf not}
LSB galaxies.

\section {Discussion}

Using the refurbished Arecibo Gregorian telescope we detected 43 galaxies
with 22.0 \mss\ \lta\ \Bmo \lta\ 25.0 \mss.  The detected galaxies range in
color from the very blue through the the first 21-cm detection of very red
LSB galaxies, while their H I mass-to-luminosity ratios range from
reasonably gas poor (M$_{HI}$/L$_B$ = 0.1 \Msol/\Lsol) to possibly the
most gas rich galaxies ever detected (M$_{HI}$/L$_B$ = 46 \Msol/\Lsol).

Analysis of the structural properties of these galaxies show a diverse
population, ranging from dwarfs to intrinsically luminous systems,
though no large, Malin I-type galaxies were identified.  We found no
correlation between the galaxies' color and M$_{HI}$/L$_B$, confirming
earlier results for HSB galaxies.  We have also discovered a contingent
of red LSB galaxies with M$_{HI}$/L$_B$ $>$ 9 \Msol/\Lsol.  These
galaxies are generally small and if the H I distribution is similar
to the optical distribution then such red galaxies are in a regime
where their is no massive star formation despite having an average
column density above the critical value.  Either star formation
is occurring without any massive star formation (e.g. the model in
Table~\ref{tab:botmod}), or the H I distribution is very extended and we have
merely detected the ``optical core'' of this extended gaseous distribution.

Finally, perhaps the most interesting discovery of this survey is the
presence of large line width galaxies which are substantially
under-luminous for their circular velocity and hence represent a
significant departure from the Tully-Fisher relation.  Moreover,
our sample shows a clear correlation
between residuals from the TF relation and gas content in the
expected sense.  Such a correlation exists primarily because most
galaxies that form the TF relation have a very similar gas mass
fraction and are in a similar evolutionary state in their history
of gas-to-star conversion.  Strong deviations away from this relation,
such as those exhibited by our sample, indicate these LSB galaxies are not
in that same state of evolution.   As much of their baryonic matter
is still in the form of gas, there should be no expectation that
these galaxies would fall in the same place on a luminosity vs
circular velocity diagram for galaxies with lower gas content, unless
the percentage of dark matter in these systems was very high.  
This in turn implies an unusual conspiracy - namely, the baryonic
mass fraction of galaxies varies with surface brightness in such
a way so as to preserve the Tully-Fisher relation.  While this
conspiracy can be resolved using MOND (e.g. McGaugh \& de Blok 1998),
it is also possible to achieve a universal baryonic mass fraction
using the kind of stellar population model here, which does
adequately account for the simultaneous existence of both
gas-rich blue and red LSB galaxies.   Under that model, we expect
very red gas-rich LSB galaxies to strongly deviate from the Tully-Fisher
relation in the sense observed here.  The next step in unraveling
all of this requires detailed investigations of the actual gas 
distributions in these red, gas-rich LSB galaxies.

\acknowledgements
Many thanks to Mike Davis for all his help and patience in
determining the gain/temperature zenith angle corrections for the L-wide 
receiver.
This research made use of NED, the NASA
Extragalactic Database.  Work on LSB galaxies at the University of Oregon
is supported by NSF grant AST 9617011.

\clearpage

\centerline{\bf TABLES}

Table~\ref{tab:obs}. The galaxy observations taken during this survey.
 
Table~\ref{tab:data}. The observed and derived data from this survey.

Table~\ref{tab:nondet}.  Limiting values for the galaxy non-detections in this survey.

Table~\ref{tab:botmod}.  The ON and OFF states of the extreme star formation model.

\clearpage

\centerline{\bf FIGURES}

\figcaption[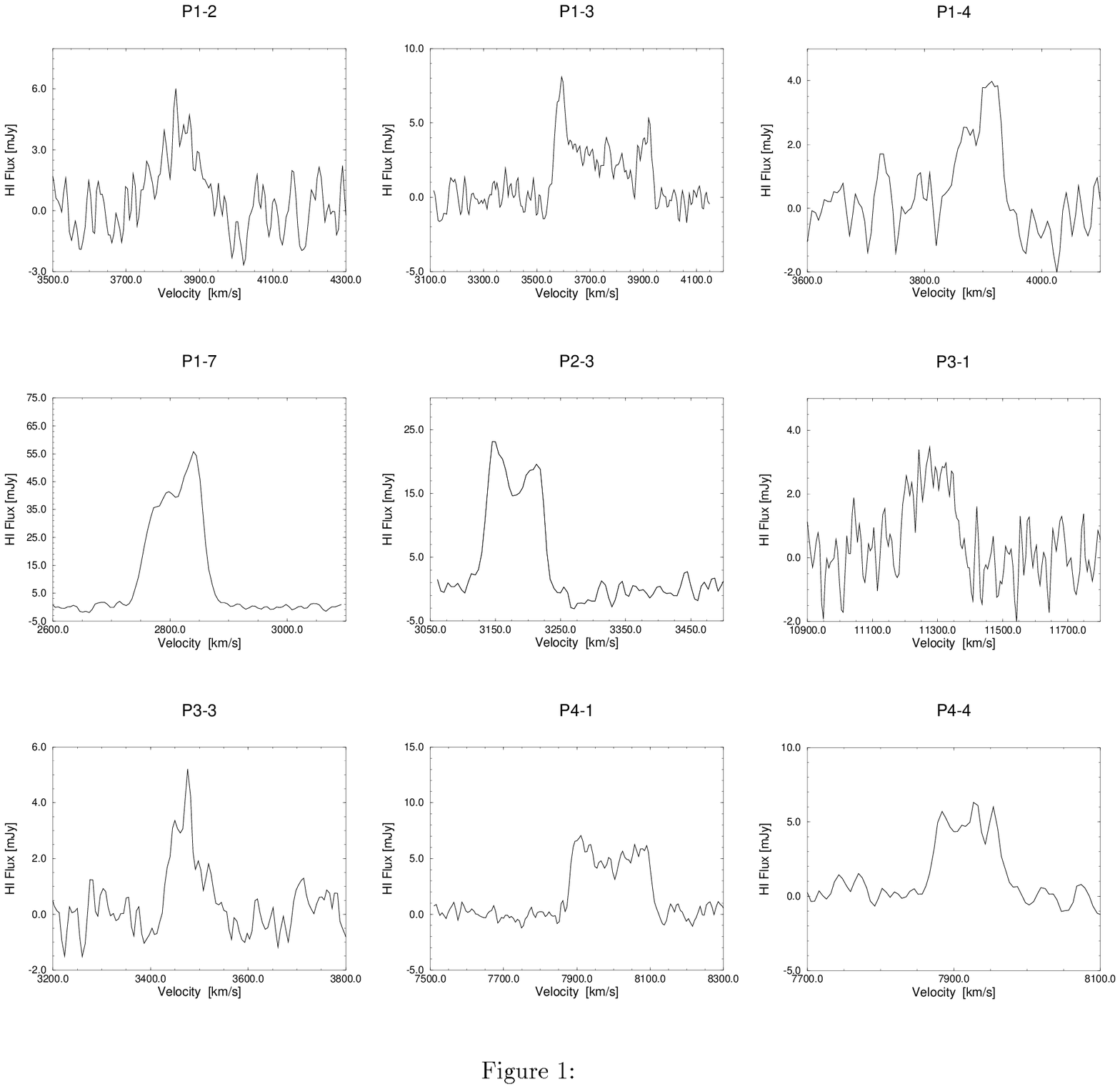]{H I profiles of all the galaxies listed in
Table~\ref{tab:data}. \label{fig:HIprof}}

\figcaption[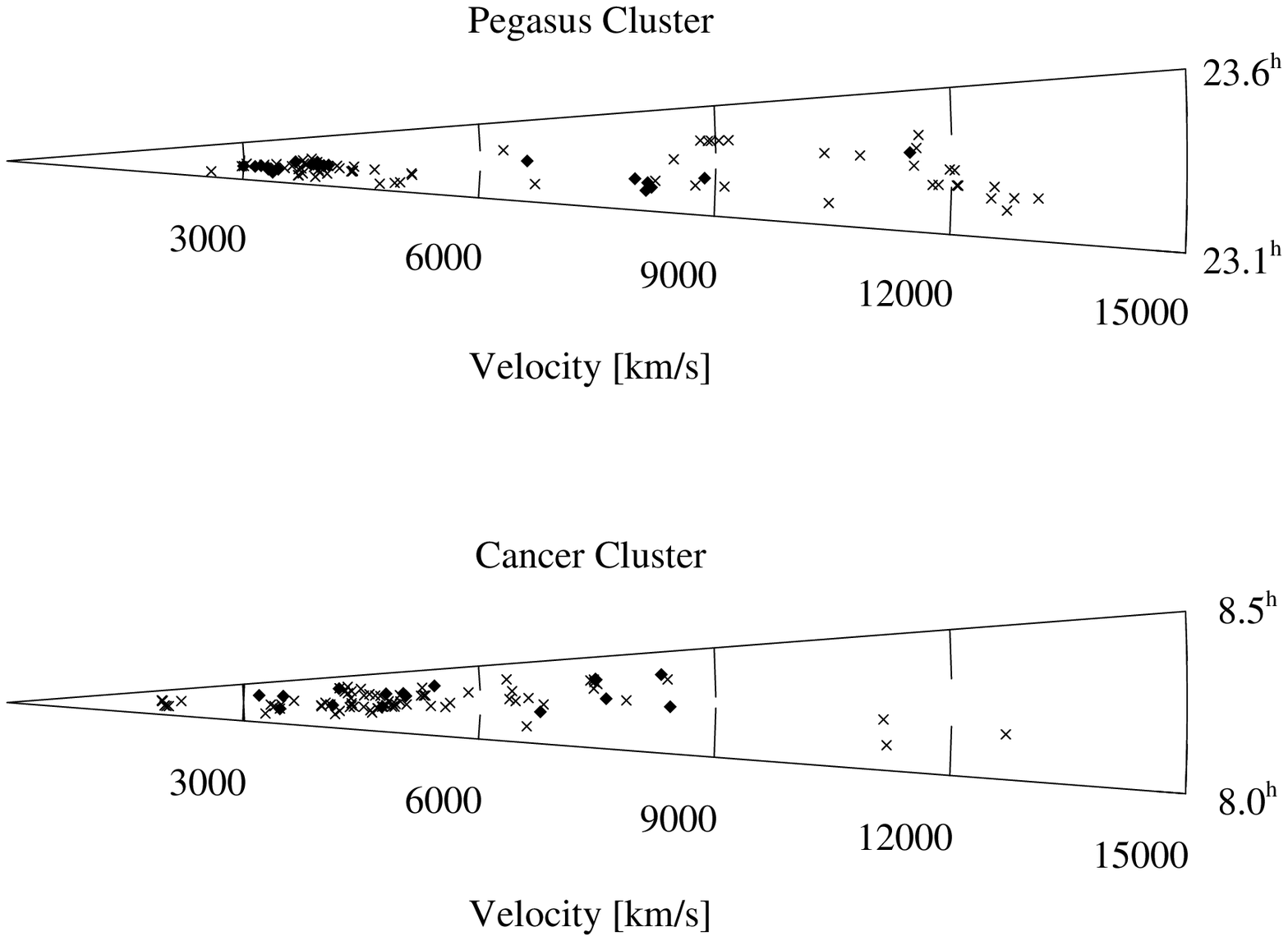]{Two dimensional projection of all the galaxies both in our
survey ($\bullet$) as well as all other galaxies with published velocities and lying in
the direction of the Pegasus ({\it a}) and Cancer ({\it b}) clusters (+) as determined from 
NED.  The plots show heliocentric velocity (in km/s) versus right ascension (B1950 coordinates).
\label{fig:rav}}

\figcaption[obs.fig3.ps]{M$_B$ versus the total H I mass ({\it a}) and the 
B$-$V color ({\it b}) for all the galaxies in this survey ($\bullet$), the LSB 
galaxies sample of de Blok, \etal (1995, 1996) (+), the dwarf spiral sample of
Schombert, \etal (1995) (X), the late-type spiral
galaxies from the Matthews, \etal survey (1997) ($\Box$), the spiral galaxies of the
Bothun, \etal survey (1982) ($\triangle$), and the Becker, \etal (1988) survey of
bright galaxies ($\diamondsuit$).\label{fig:Bcomp}}

\figcaption[obs.fig4.ps]{Comparisons of M$_B$ versus central surface brightness ({\it a})
and scale length ({\it b}) for all the galaxies in this survey.  \label{fig:morph}}

\figcaption[obs.fig5.ps]{Total integrated H I mass ({\it a}) and M$_{HI}$/L$_B$ ({\it b})
versus B$-$V for all the galaxies in detected with our survey ($\bullet$), the LSB
galaxies sample of de Blok, \etal (1995, 1996) (+), the dwarf spiral sample of
Schombert, \etal (1995) (X), the late-type spiral
galaxies from the Matthews, \etal survey (1997) ($\Box$), the spiral galaxies of the
Bothun, \etal survey (1982) ($\triangle$), and the Becker, \etal (1988) survey of
bright galaxies ($\diamondsuit$). \label{fig:red}}

\figcaption[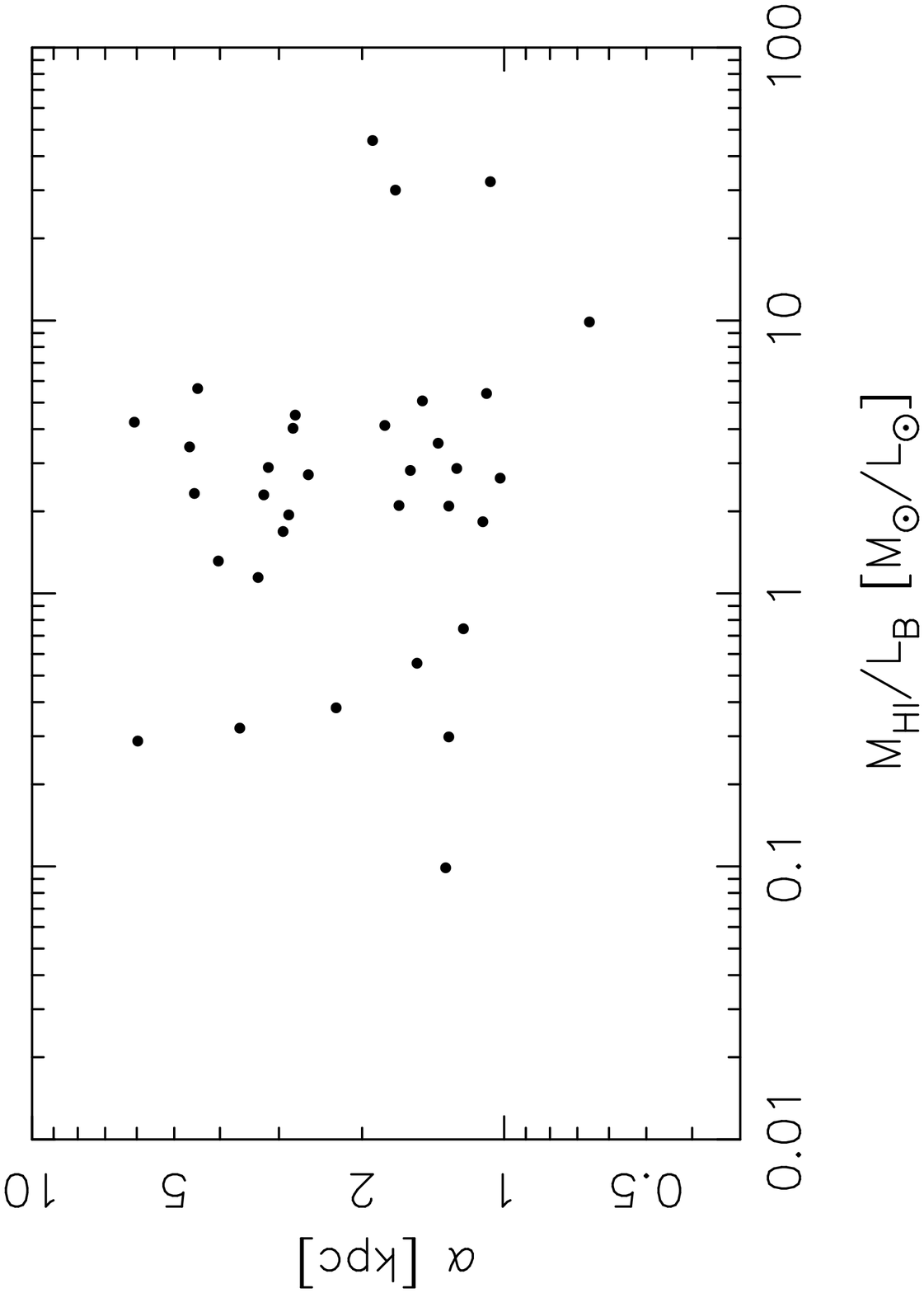]{M$_{HI}$/L$_B$ versus scale length (in kpc)
for all the galaxies in detected with our survey. \label{fig:MLBalp}}
 
\figcaption[obs.fig7.ps]{Corrected ({\it a}) and uncorrected 50\% ({\it b})
velocity widths versus absolute B magnitude for all the galaxies in detected with our survey.
The solid and dashed lines are the 1$\sigma$ and 2$\sigma$ fits to the
Tully-Fisher relation by Zwaan, \etal (1985).\label{fig:tf}}

\figcaption[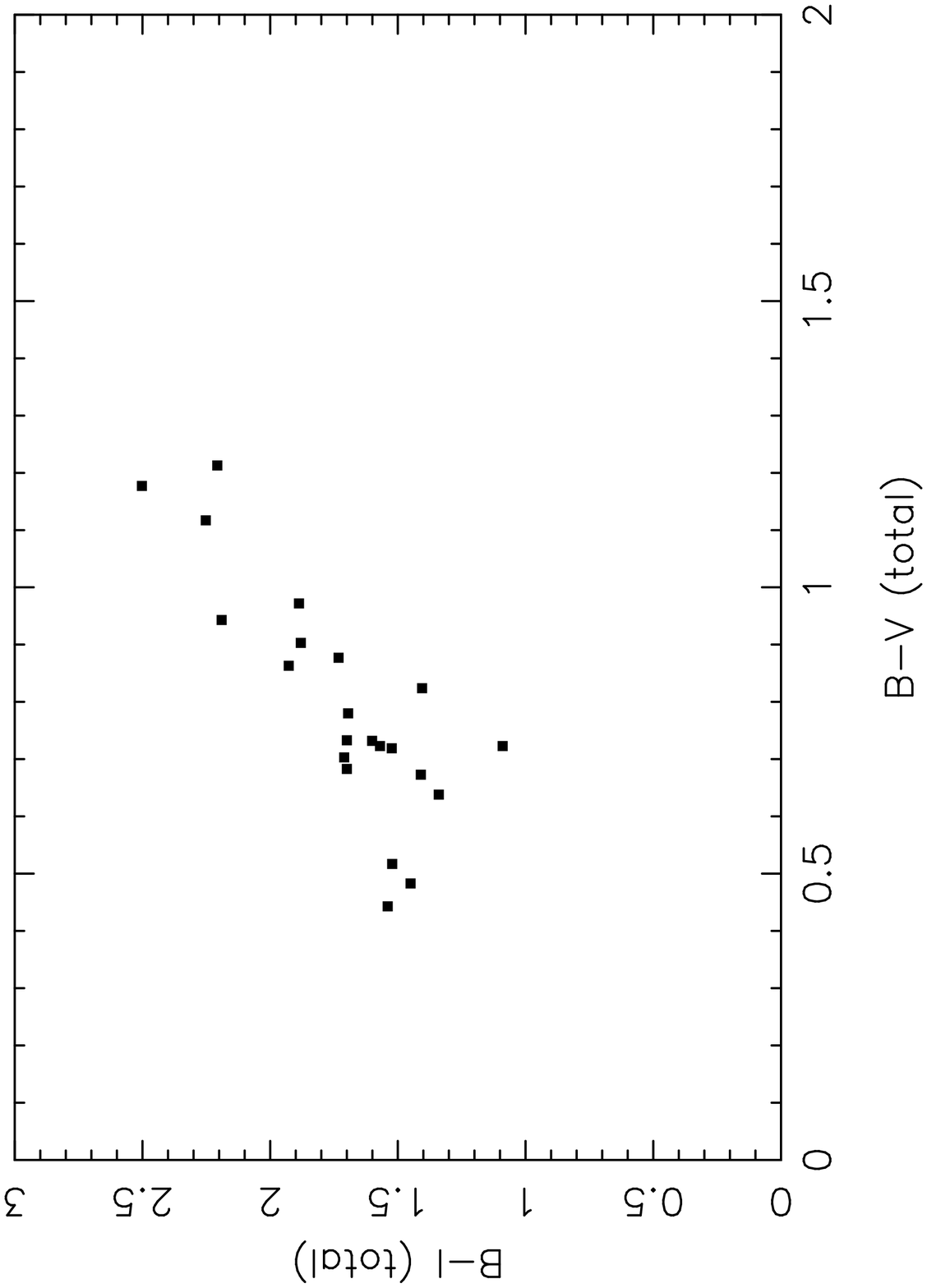]{B$-$V versus B$-$I for all the galaxies detected with
our survey. \label{fig:bvbi}}

\figcaption[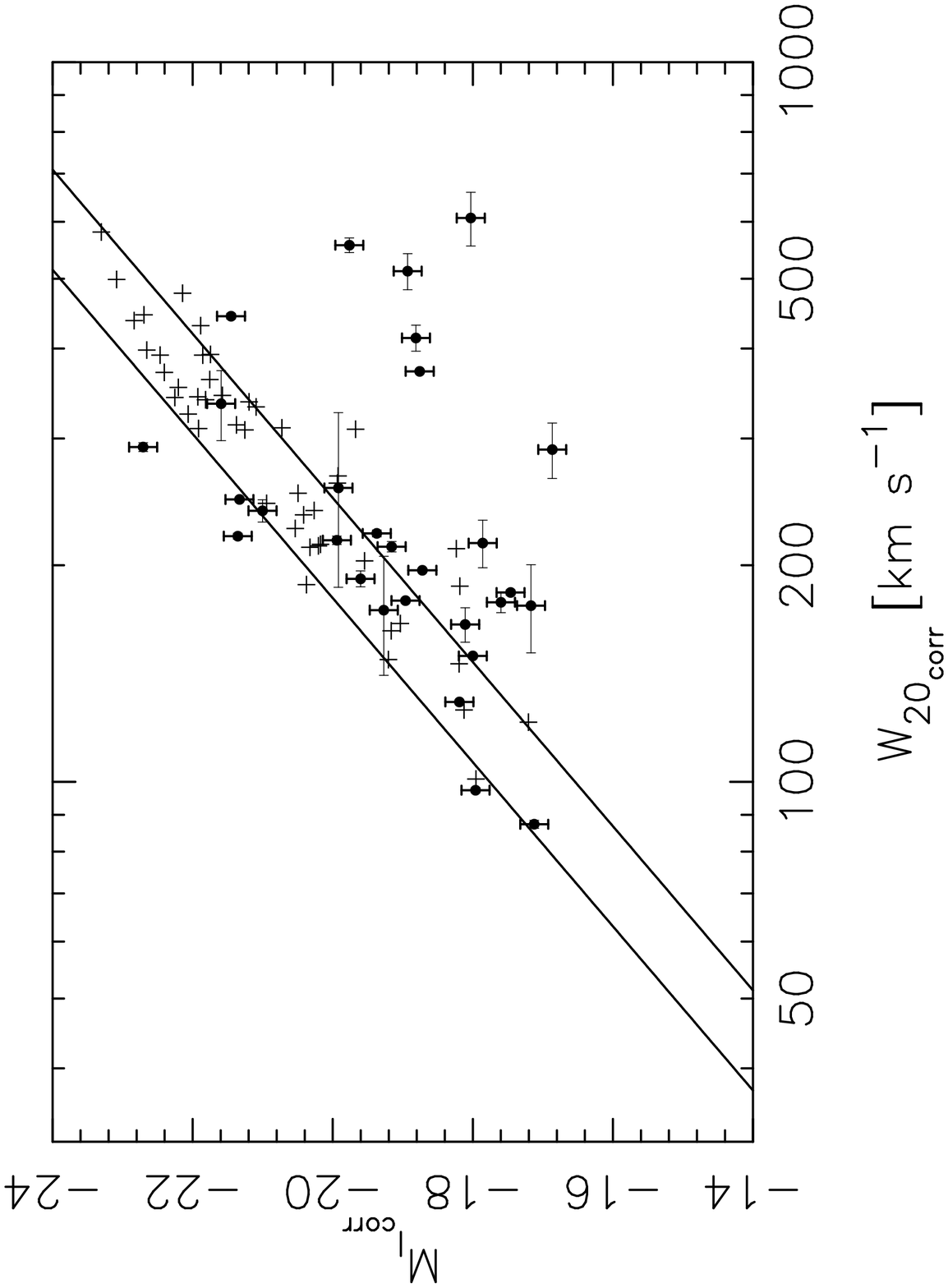]{Corrected  50\% velocity widths versus absolute I 
magnitude for all the galaxies in detected with our survey ($\bullet$).  The solid lines
are the 1$\sigma$ fits to the Virgo data of Pierce \& Tully (1988) (+), using 
a Virgo distance of 20h$_{75}^{-1}$ Mpc.  \label{fig:tfi}}

\figcaption[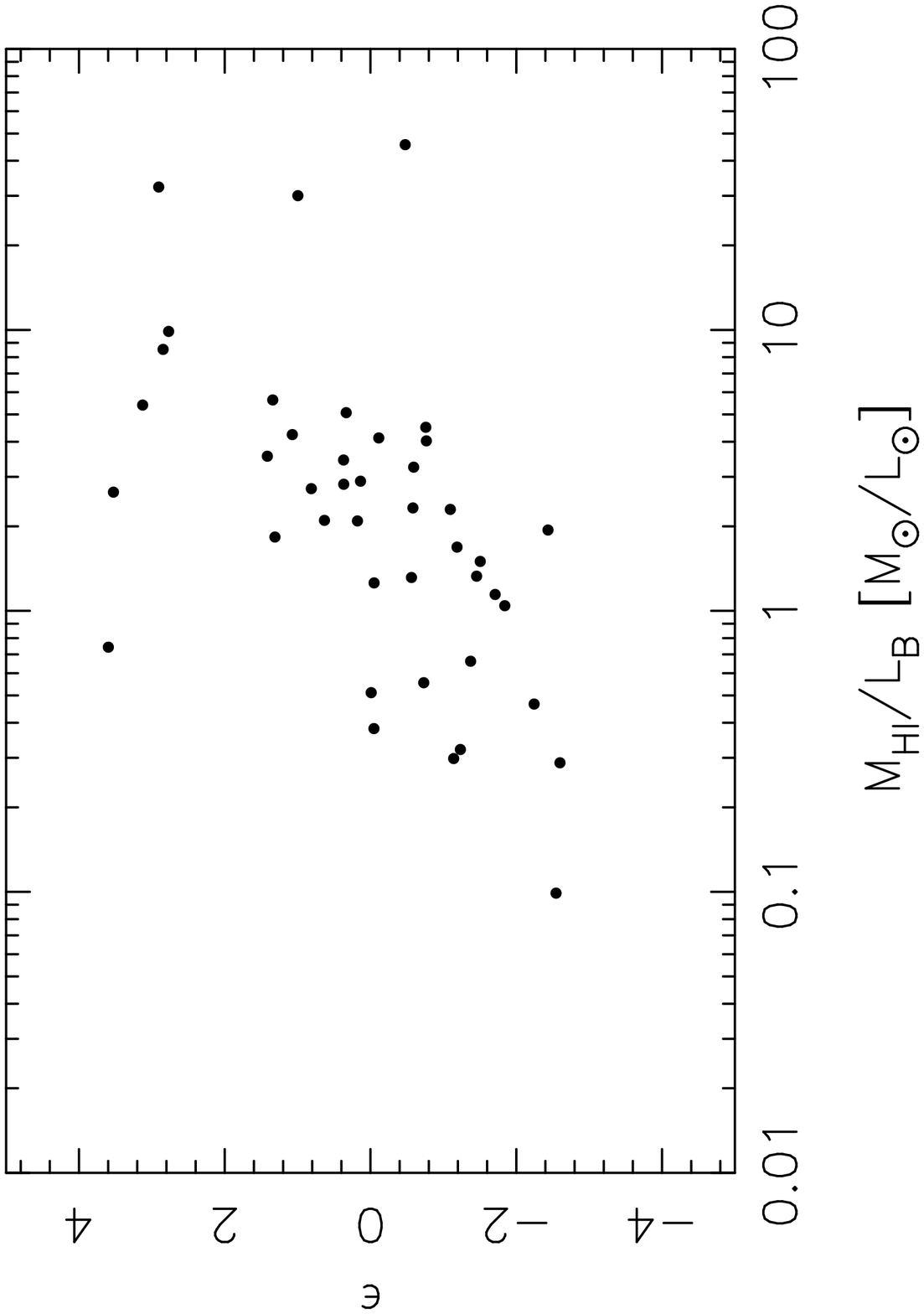]{Residuals from fitting the galaxies detected with our survey
to the Tully-Fisher relation, versus M$_{HI}$/L$_B$. \label{fig:resid}}

\clearpage

\begin{table}
\dummytable{\label{tab:obs}}
\end{table}

\begin{table}
\dummytable{\label{tab:data}}
\end{table}

\begin{table}
\dummytable{\label{tab:nondet}}
\end{table}

\begin{table}
\dummytable{\label{tab:botmod}}
\end{table}

\clearpage

\end{document}